# Mobile Learning Game Authoring Tools: Assessment, Synthesis and Proposals

Aous Karoui[1], Iza Marfisi-Schottman[1], Sébastien George[1]

[1] COMUE UBL, Université du Maine, EA 4023, LIUM, 72085 Le Mans, France
{Aous.Karoui,Iza.Marfisi,Sebastien.George}@univ-lemans.fr

**Abstract.** Mobile Learning Games (MLGs) show great potential for increasing engagement, creativity and authentic learning. Yet, despite their great potential for education, the use of MLGs by teachers, remains limited. This is partly due to the fact that MLGs are often designed to match a specific learning context, and thus cannot be reusable for other contexts. Therefore, researchers have recently designed various types of MLG authoring tools. However, these authoring tools are not always adapted to non-computer-scientists or non-game-designers. Hence, we propose in this paper to focus on five existing MLG authoring tools, in order to assess their features and usability with the help of five teachers, who are used to organizing educational field trips. In the second part of this paper, we present an approach for designing a MLG authoring tool, based on the lacks identified through the analysis, and tailored to the teachers' different profiles and needs.

## 1 Introduction

Mobile Learning Games (MLGs) have proven their efficiency not only for improving students' engagement but also for improving effective learning in certain studies. For example, *Frequency1550* [1], is a MLG designed to learn about medieval Amsterdam History, which helped high-school students get higher scores on the knowledge test than with regular lessons. Other MLGs have also proven their effectiveness for improving engagement (e.g. *TheMobileGame*, designed to introduce a university campus in Berlin to new comers, that students preferred to the classic visiting tour [2]) and creativity (e.g. *skattjakt*, a MLG co-designed with students to promote physical activity while learning a novel [3]).

Now that smartphones are widespread among teenagers and that schools are increasingly equipped with tablets [4], using MLGs in class has never been simpler. Moreover, MLGs can take advantage of mobile devices' assets to enhance learning and gaming experience (e.g. position, orientation and proximity sensors, media capturing and recording, augmented reality on learning objects…). However, very few teachers actually create a MLG for their course. Therefore, we propose in this paper, to analyze the usability and the features offered by current MLG authoring tools, in order to understand this phenomenon. As a second step, we present our work for designing a MLG authoring tool based on the lacks identified through this analysis and tailored to the teachers' different profiles.

## 2   Methods

### 2.1   Screening Set

To outline our screening set, we define the MLGs that we are interested in, as following: a MLG, is a mobile app, combining pedagogic content with a playful scenario to enhance learning, and taking advantage of mobile devices' assets (e.g. location, orientation and proximity sensors, media capturing and recording, augmented reality…).

Consequently, we focused on authoring tools that could be used to create these MLGs, whether they were initially intended to produce MLGs, mobile games or even general mobile apps. As a matter of fact, many interesting Serious Game authoring tools found in literature could not be taken into account in this study, as they do not provide the mobile assets nor do they make mobile apps.

### 2.2   Selection Method

We chose to focus only on functional authoring tools available on the Internet and usable by teachers who do not have any programming skills or game-design experience. Many MLG authoring tools we found in literature could not be a part of this analysis, since they were still under development or not accessible for public use[1]. Other interesting tools such as ARLearn could not be included since they were not destined to be used by teacher on their own [5]. The five authoring tools we selected are freely accessible online. They were selected on account of provided technical features, essentially the mobile assets needed to create efficient MLGs that we have identified in a previous work [6]. In addition, we took into account the provided assistance to educational and gameplay design (i.e. setting up learning content and provided game mechanics).

### 2.3   Assessment Method

Our authoring tool assessment is based on a criteria grid[2] regarding two perspectives:

**The provided technical features:** We identified the features provided by each authoring tool by trying to reproduce existing MLGs that we identified in previous work [6] which proved to enhance learning and engagement. A symbolic score concludes each assessment, in order to obtain comparable results between the five authoring tools.

**The authoring tools usability:** In order to measure usability, we firstly used Bastien & Scapin's guidelines for measuring Human-Computer Interfaces (HCI) usability [7]. We then asked five teachers, who organize educational field trips every year, to try to design one of their usual outing activities, with these authoring tools, while adding a few game mechanics that we recommended such as scores and timers. This time, the assessment score was the average between our rating based on the ergonomic criteria cited above and the teachers rating based on their user experience.

---
[1] http://perso.univ-lemans.fr/~akaroui/oa_list.htm
[2] The assessment grid is available here: http://perso.univ-lemans.fr/~akaroui/ot_grid.htm

## 3 Authoring Tool Analysis

### 3.1 ARIS (Augmented Reality and Interactive Storytelling)

The ARIS[3] project started in 2008 at the University of Wisconsin, in order to design an open source tool for creating learning games for iPhones [8].

**Feature Assessment.** Technically, ARIS incorporates geolocation, QR-codes, uploading media content (photos, sounds), and options for managing teams. The web editor provides a variety of "games objects" to interactively include media content into scenarios. Those "game objects" are created by the designer and then placed on the map representing the game field. For example, "*plaques*" consist in textual tags for showing static information to players. They can be used for tours and for providing narrative events in a game. Additionally, the "*conversation*" object is another way of providing information to players. They combine text and media resources to provide virtual conversations for players to facilitate their progression in the game. ARIS also provides "*quests*" which comprise a notification system to help players focus on what they can and should be doing. In addition to the "game objects", ARIS put to use "*locks*" components, which are triggers defining how players access content by turning the "game objects" visible or invisible during the game. Finally, in order to create coherent game steps, all elements should be held together within "*scenes*" which are abstract units organizing triggers and game objects. Furthermore, ARIS provides a JavaScript API that allows programmers to modify the MLGs in detail. These customizations range from adding interactive mini-games, to complex menu structuring, navigation flow redirection and altering a player's inventory in ways not currently supported by ARIS.

ARIS is surrounded by a large community of users and developers who continuously contribute to the project by adding new features and animating online forums. Considering all the features cited above, we assigned 4/5 to ARIS for its features.

**Usability Assessment.** From a usability perspective, the major drawback of ARIS is the unusual vocabulary related to the "game objects" presented in the previous paragraph. Indeed, the five teachers who tested the authoring tools with us, found that terms such as "*plaques, quests, conversations …*" are completely incomprehensible. Using ARIS is therefore impossible without consulting the online manual and the tutorials that need several hours to be discerned. Three of the five interviewed teachers found that tutorials were "too long" and said that they would have "given up". Finally, the feature customization part is reserved for programmers as it relies on the use of JavaScript programming language. Thus, considering ergonomic criteria and the teachers' feedback for the usability test, we assign 2/5 to ARIS.

---

[3] http://arisgames.org/make/

### 3.2 App Inventor 2

App Inventor *2* is an authoring environment for Android mobile apps[4]. It is also very useful for rapid prototyping mobile serious games [9].

**Feature Assessment.** The App Inventor 2 web editor offers a wide range of pallets, from primary mobile apps components (e.g. buttons, labels, sliders …), to elaborate data storage components (e.g. storing files, data tables, database …). Sensors (e.g. position, orientation and proximity sensors), multimedia and connectivity tools (e.g. Bluetooth, SMS, web connectivity …) are provided as well. All the items can be knit together thanks to the *blockly*[5] library, incorporated within the App Inventor 2 editor. Indeed, *blockly* is a powerful block programming interface allowing users with a low programming background (e.g. children, programming novices) to easily link and configure items in order to get a functional program. These items could be used as dependencies and triggers for MLG design.

App Inventor 2 is a widely used authoring tool. A large community of designers but also developers, contribute to its content enrichment every day. Considering all the features cited above, we attribute 4.5/5 to App Inventor 2 for its features.

**Usability Assessment.** App Inventor 2 provides a rich Graphical User Interface (GUI) based on the SPI (*Single Page Interface*) model. Indeed, all the design components are available on the main design page, categorized by type and attainable by *drag-and-drop*. Although, the main design page contains several boxes (e.g. palette, items properties, screen viewer …) and two main views. The first one is intended for components set up to the mobile viewer box. The second view is intended to coordinate components in order to get a working program via the *blockly* editor. However, this design way, even though much simpler than real programming, is not intended for people without programming background. Indeed, the teachers we interviewed had not any technical background and gave us feedbacks such as "this is for computer-scientists", 'I cannot go through it". Consequently, App Inventor 2 usability score was set to 1.5/5.

### 3.3 Pocket Code

Pocket Code[6] is an open source authoring tool realized within the Catrobat project [10] for creating and sharing mobile learning apps by children and teenagers.

**Feature Assessment.** Pocket Code is based on three main components (*scripts, graphics* and *sounds*) that could be highly customized and linked together in order to create playful scenarios. This coordination is feasible by assembling visual programming blocs as same as with App Inventor 2. Thus, Pocket Code incorporates QR-code set up, multimedia content managing (i.e. text, image and sound) and several types of sensors (i.e. location, orientation and proximity). However, Pocket Code is entirely executable on mobile devices and then allows designers to create mobile apps even on their smartphones.

---

[4] http://appinventor.mit.edu/explore/
[5] https://developers.google.com/blockly/
[6] http://www.catrobat.org/intro/

The editor also includes a game scene recorder to easily share created scenes on YouTube. All the projects created by users are open-source and available online in order to be reusable. For example, interesting education-specific resources have been created by the Technology Enhanced Learning (TEL) community and are available online[7]. Pocket Code's large users and developers' community keeps empowering tutorials and creating useful Frameworks such as "Pocket Paint"; a library that enables Pocket Code users to edit images. Hence, Pocket Code's features score was set to 4/5.

**Usability Assessment.** Although setting the design process on mobile is an original feature for taking advantage of mobiles portability, we believe that this design mode considerably limits the ergonomic comfort required to create MLGs. Yet, the minimized screen size does not provide a full vision of the scenario components (i.e. *scripts, graphics,* and *sounds*). Even these components can be shown on several tabs, a non-complete items view implies memorizing too much data and then would considerably augment memory load especially for novices designers. Indeed, Bastien & Scapin [7] report that this would absolutely worsen the user experience. Besides, some teachers clearly said that they don't prefer designing on mobile as they do not have access to their educational resources, typically stored on their computers. Other difficulties may arise when designing on mobile, such as problems with inaccuracy of touch interactions. Consequently, the average score for Pocket Code for usability is 1.5/5.

### 3.4 FURET FACTORY

Furet Factory[8] is an online platform for designing mobile games. It was developed by *Furet Company*[9], specialized in designing cultural heritage games.

**Feature Assessment.** Several types of games are available (e.g. treasure hunt, interactive tour, quizz). The game stages can also be set up by customizing the challenges: puzzle, multiple choice question, riddle, geolocation. The points earned by players translate into levels of expertise (e.g. Amateur Detective, Chief Inspector, Emeritus Adventurer, etc.). In addition, players can also evaluate the games and assign points to game designers. Score tables are published online and players can invite their friends to play via social networks.

The technical features for Furet Factory are very limited in comparison to the authoring tools analyzed above. Indeed, it lacks features such as QR-code support, including rich multimedia items (e.g sounds, videos) and configurable triggers. Moreover, it does not handle multiplayer games or provide means of communication between players. Consequently, we attribute 1.5/5 to Furet Factory, for its features.

---

[7] https://edu.catrob.at/
[8] http://www.furetfactory.com/
[9] http://www.furetcompany.com/

**Usability Assessment.** The design process proposed by Furet Factory is instantaneously apprehended. The notions introduced for game items and steps are fully clear and make the design process intuitive even for a first-time user. In consequence, there is no need to go through tutorials to use this authoring tool. In terms of guidance, components information is shown on demand and through pop-up windows in an interactive way. All the teachers participating in the test where comfortable with Furet Factory and gave us positive feedback. In consequent, the average between the resulting score from the ergonomic criteria and the teachers' usability rating is 4/5.

### 3.5 mLearn4web

mLearn4web[10] is an open-source authoring tool for creating mobile learning activities [11] that can be used for creating MLGs.

**Feature Assessment.** mLearn4web incorporates the essential features for taking advantage of mobile assets. Then, geolocation, multimedia content management (sounds, videos, images) and QR-code support are provided. The resulting mobile app is generated on a web responsive format, making it compatible with all mobile devices. However, as mLearn4web is not initially intended for MLGs, it does not provide items that could be set up to behave as game mechanics (e.g. scores or timers), as it is possible with App Inventor 2. Similarly, there is no way to alter the linear activities sequences. Consequently, based on a MLG design perspective, we attribute 2.5/5 for its features.

**Usability Assessment.** The simple design interface does not require specific tutorials to get familiar with. The GUI is interactive and content can be intuitively added by *drag-and-drop*. The design process consist in creating screens (which will contain activities) and gradually adding resources to them.

Although the provided components are not complex to understand, the tool doesn't provide any guidance or help on demand. For this reason, the teachers found mLearn4web less practical than *Furet Factory*, even though it provides an intuitive interface. In consequent, the average between the notation resulting from the ergonomic criteria and the teachers' usability notation is 3/5.

## 4 Synthesis

### 4.1 Analysis Summary

According to the analysis detailed above, we notice that the authoring tools which have a top rating for their features, have very low scores for their usability and vice versa. Thus, the analyzed authoring tools can be split into two categories. The first category is composed of the authoring tools that offer rich low-level-item-based GUIs, such as *ARIS*, *App Inventor2* and *Pocket Code*. Even though it is possible to create MLGs with these tools, the effort and expertise required to use them is overwhelming for teachers.

---

[10] http://www.mlearn4web.eu/

Indeed, reading the user manuals and watching the video tutorials to learn how to aggregate low-level items (e.g. text, buttons, media resources) into game mechanics (e.g. game units, scores, timers) demands a considerable effort. This effort was considered unacceptable by three of the five interviewed teachers, while the two others reported that they would prefer easier authoring tools. The second category covers the authoring tools that include few or limited features, but which are relatively simple to use, such as *Furet Factory* and *mLearn4web*. The problem here is that these authoring tools do not provide enough design features to create effective MLGs, such as those cited in the introduction. If the authoring tools in this second category provided more features, would this be the solution? According to HCI specialists, augmenting information density in general, implies augmenting perceptive and cognitive workload [7]. Therefore, we believe that augmenting authoring tools features would make them join the first category and so the usability problem would persist.

### 4.2 Understanding the Teachers Needs

To explore the previously discussed issues, we sent an online questionnaire to several teachers' mailing lists asking if the teachers would like to try MLGs during their learning outings. Out of the 26 teachers who responded, we selected five teachers to conduct qualitative interviews. We selected these teachers in such a way to have a variety of teaching levels (i.e. middle-school, high-school and college) and field trips (i.e. analyzing landscapes (botany), examining rocks (geology), and observing biodiversity (biology)). Each interview lasted between one and two hours, and consisted in testing a couple of the authoring tools analyzed above. Then we asked the teachers send us their feedback and usability scores for the remaining authoring tools by email. The second part of the interview consisted in co-designing the GUI of a MLG authoring tool that would match their needs (discussed in the last subsection of this paper).

Three of the five interviewed teachers affirmed that they were interested in creating MLGs if it did not take them more than half a day. Actually, they reported that, to start with, they just want to reproduce the pedagogic content of their field trips on mobile devices, and add some game mechanics (i.e. scores, timers, collaboration) to create a playful scenario. The two other teachers affirmed that they already had some experience in learning games design and would probably spend more than half a day in the design process. The analysis of the questionnaire sent out to teachers' mailing also shows this disparity. Out of the 20 answers for this question, 14 teachers said that they were willing to try MLGs and 6 said that it would depend on the required investment level.

Moreover, we do not exclude the fact that teachers' engagement in designing MLGs could vary, depending on their growing experience and also on the authoring tool usability. Thus, this could imply changing in the teachers' investment and have to be taken into account in the design approach that we propose.

Following the five interviews, we notice that teachers are initially divided in two categories. The first one comprises teachers who do not have any game design experience but are quite interested in the topic and would like to try MLG creation, if it doesn't take too much time to be set up. The second category comprises teachers who are motivated for using MLGs and would be willing to put in more effort, if they can create the MLGs they want. In the next section, we propose an innovative approach for satisfying the needs of different teachers' profiles. Indeed, even if we detect two main users' profiles, it could be seen as a continuum and intermediate profiles could exist.

## 5     Current Work and Proposals

### 5.1    Authoring Tool Complexity in TEL

Since research on MLG authoring tool usability is lacking, we decided to look for solutions for the authoring tool usability problem in larger areas such as TEL. Even though most of research in TEL focuses on learning environments' usability, there are several works about the usability of learning environment authoring tools. Murray for example, summarized in 2004 [12], the authoring tool design tradeoffs in three categories: power, usability and cost. He proposes a collaborative design with multiple roles as the optimal solution for the authoring tool complexity problem. Given that in this case, the authoring tool would be powerful as each of the designing team members would contribute to the authoring process, and usable as each participant would not face difficulty in handling his/her own part of authoring. In the same context, Ritter [13] adheres totally to the idea that the authoring process should be performed by a designing team. Furthermore, he suggests that different interfaces should be built to support different roles within the designing team "rather than having one huge monolithic authoring tool". Similarly, Oja, in a study for improving usability in complex software systems [14], concludes that systems' interfaces should anticipate the variety of roles and areas of expertise.

Nevertheless, we do not embrace the idea of collaborative design as it implies hiring costly authoring experts. Besides, in his latest research, Murray [15] characterizes the complexity of systems in general terms (such as Complexity Science and Hierarchical Complexity Theory) and updates the tradeoffs that were presented in [12]. In [15], the idea of the collaborative authoring tool does not seem to be retained anymore as it is not brought into discussion again.

Even though, we retain the idea of differentiating interfaces, not for different roles but for the different teachers profiles identified on the previous section. Furthermore, it could be perfectly associated to the Hidden complexity quoted by Murray in [12]. Indeed, the Hidden complexity is a strategy for making tools more usable by hiding the advanced tools and making common and easy tools more salient. We explain our approach of using those insights in the next subsection.

### 5.2    A Multi-View GUI Based on a Nested Design Process

Based on the teachers' interviews, and to deal with the diversified needs we highlighted in subsection 4.2, we aim to design an authoring tool with a multi-view GUI. The several views would not be intended to different roles, as in collaborative design, but rather to match the teachers' various levels of expertise. Because the authoring task requires the ability to conceptualize and structure the concepts from a high level as explained by Murray [15], we propose an authoring tool comprising mainly of three views:

1. A "Standard view" providing a couple of object types that can be slightly adjusted (e.g. gps coordinates of points of interest (POI), learning and questions content). This view will allow the first category of teachers (cited above) to rapidly design a basic playful scenario with preconfigured game mechanics (e.g. a linear game unit order, a standard way of counting scores).
2. An "Intermediate view" allowing designers to go further in details, in order to better adjust their scenarios. This time, the teachers can configure the score mechanisms, the radius of POI, game unit triggers and dependencies …. This view is intended to the second category of teachers (cited above).
3. An "Expert view" allowing the most expert designers to go even further in details. We aim to provide custom component creation at this level and programming features to create the logic between them.

From a conceptual perspective, the underlying data model of our authoring tool is based on mapping high-level components, which are comprehensible by teachers, such as points of interests, activities and clues to low-level executable components (e.g. multimedia resources, buttons, textual items …). From the design process perspective, we intend to provide a nested design process, meaning that views are embedded in each other according to the Hidden complexity theory. The content to be shown in the previously presented three views, was decided by consulting the five teachers. Thus, every view leads to the other as if one chooses to navigate from "standard" to "intermediate", looking for more options to set up. Likewise, navigation in the opposite is necessary if one doesn't feel comfortable with the "intermediate" or the "expert" view.

The three views discussed above have been co-designed with the five interviewed teachers on graphical mockups. Even though, we decided to begin with three levels, this number is not definitive and surely can be adjusted according to intended users, especially if we generalize the use of this approach outside the MLG design field. Then, our next step is to test a first MLG authoring tool prototype, implementing the insights discussed above, with the teachers who answered our online questionnaire.

## 6    Conclusion and perspectives

In this paper we analyzed five authoring tools that can be used to create Mobile Learning Games (MLGs). This study identifies the reasons that are slowing down the use of MLGs by teachers, despite the material resources available and the MLGs' potential for learning. Our analysis consisted in assessing the technical features provided by each authoring tool that we tested by reproducing existing MLG scenarios. The second part of the analysis consisted in assessing the usability of each authoring tool, based on a HCI usability criteria and the feedback provided by five teachers organizing educational field trips.

In the second part of this paper, we presented the main issues that explain why these authoring tools are not used by teachers: either they offer very rich functionalities but are very complicated to use, either they are simple to use but do not offer the necessary functionalities to design MLGs. We therefore propose our approach of a MLG multi-view authoring tool, based on a nested design process. We are currently collaborating with the five teachers to design the mock-up models of three different interfaces: a standard view, an intermediate view and an expert view, which gradually show more and more functionalities.

More generally, authoring tool usability is a persistent problem in the TEL field. As a consequence, our future work will also be focused on generalizing the multi-view model, based on the nested design approach, to TEL systems.

## References

1. Admiraal, W., Huizenga, J., Akkerman, S., Dam, G. ten: The concept of flow in collaborative game-based learning. Comput. Hum. Behav. 27, 1185–1194 (2011).
2. Schwabe, G., Göth, C.: Mobile learning with a mobile game: design and motivational effects. J. Comput. Assist. Learn. 21, 204–216 (2005).
3. Spikol, D.: Exploring Novel Learning Practices Through Co-Designing Mobile Games. Presented at the (2009).
4. Johnson, L., Adams Becker, S., Estrada, V., Freeman, A., Kampylis, P., Vuorikari, R., Punie, Y.: Horizon Report Europe: 2014 Schools Edition. New Media Consort. (2014).
5. Klemke, R., van Rosmalen, P., Ternier, S., Westera, W.: Keep It Simple: Lowering the Barrier for Authoring Serious Games. Simul. Gaming. 46, 40–67 (2015).
6. Karoui, A., Marfisi-Schottman, I., George, S.: Towards an Efficient Mobile Learning Games Design Model. In: European Conference on Games Based Learning. pp. 276–285. , Steinkjer, Norway (2015).
7. Bastien, J.M.C., Scapin, D.L.: Ergonomic criteria for the evaluation of human-computer interfaces. INRIA (1993).
8. Gagnon, D.J.: ARIS. The University of Wisconsin-Madison (2010).
9. Rouillard, J., Serna, A., David, B., Chalon, R.: Rapid Prototyping for Mobile Serious Games. In: Learning and Collaboration Technologies. Technology-Rich Environments for Learning and Collaboration. pp. 194–205. Springer International Publishing (2014).
10. Slany, W.: Pocket Code: A Scratch-like Integrated Development Environment for Your Phone. In: Proceedings of the Companion Publication of the 2014 ACM SIGPLAN Conference on Systems, Programming, and Applications: Software for Humanity. pp. 35–36. ACM, New York, NY, USA (2014).
11. Zbick, J., Nake, I., Jansen, M., Milrad, M.: mLearn4Web: A Web-based Framework to Design and Deploy Cross-platform Mobile Applications. In: Proceedings of the 13th International Conference on Mobile and Ubiquitous Multimedia. ACM, New York, USA (2014).
12. Murray, T.: Design tradeoffs in usability and power for advanced educational software authoring tools.-SADDLE BROOK THEN ENGLEWOOD CLIFFS NJ-. 44, 10–16 (2004).
13. Ritter, S., Blessing, S.B., Wheeler, L.: Authoring Tools for Component-Based Learning Environments. In: Murray, T., Blessing, S.B., and Ainsworth, S. (eds.) Authoring Tools for Advanced Technology Learning Environments. pp. 467–489. Springer Netherlands (2003).
14. Oja, M.-K.: Designing for Collaboration: Improving Usability of Complex Software Systems. In: CHI '10 Extended Abstracts on Human Factors in Computing Systems. pp. 3799–3804. ACM, New York, NY, USA (2010).
15. Murray, T.: Coordinating the Complexity of Tools, Tasks, and Users: On Theory-based Approaches to Authoring Tool Usability. Int. J. Artif. Intell. Educ. 26, 37–71 (2015).